\documentclass[preprint]{article}
\usepackage{authblk}
\usepackage{graphicx}
\usepackage{dcolumn}
\usepackage{bm}
\usepackage{amsmath}
\usepackage{amssymb}
\usepackage{multirow}
\usepackage{caption}
\usepackage{subcaption}
\usepackage{epstopdf}
\usepackage{hyperref}

\title{\bf Behaviours of rip cosmological models  in $f(Q,C)$ gravity}
\author[1]{Amit Samaddar\thanks{samaddaramit4@gmail.com}}
\author[2]{S. Surendra Singh\thanks{ssuren.mu@gmail.com}}
\author[3]{Shah Muhammad\thanks{skabeer@ksu.edu.sa}}
\author[4,5]{Euaggelos E. Zotos\thanks{evzotos@physics.auth.gr}}
\affil[1,2]{Department of Mathematics, National Institute of Technology Manipur, Imphal-795004,India.}
\affil[3]{Department of Mathematics, College of Science, King Saud University, P.O. Box 2455, Riyadh 11451, Saudi Arabia.}
\affil[4]{Department of Physics, School of Science, Aristotle University of Thessaloniki, GR-541 24, Thessaloniki, Greece.}
\affil[5]{S.M. Nikolskii Mathematical Institute of the Peoples' Friendship University of Russia (RUDN University), Moscow, 117198, Russia.}
\begin{document}

\maketitle

 \textbf{Abstract}: In this study, the Universe's rip cosmology theories have been provided for the $f(Q,C)$ gravity theory, where $Q$ and $C$ stand for the non-metricity scalar and boundary term. We assumed $f(Q,C)=\alpha Q^{n}+\beta C$ and analyzed the nature of the physical parameters for the Little Rip, Big Rip and Pseudo Rip models. In the LR and PR models, the EoS parameter exhibits phantom characteristics but remains closely aligned with the $\Lambda$CDM line. After investigating the energy conditions, we recognised that our model violates the strong energy constraint. Avoiding singularity situations has been noted in all of these accelerated models. The characteristics of the jerk and snap parameters have been investigated. Our model provides an effective description of the Universe's evolutionary history and fits well with contemporary cosmic data.\\
 
 
 \section{Introduction}\label{sec1}
 \hspace{0.5cm} Supernova Type Ia (SNeIa), Cosmic Microwave Background (CMB) and Baryon Acoustic Oscillations (BAO) observations demonstrate the Universe's accelerating expansion \cite{Misner73,Clifton12,Riess19}. In the context of General Relativity (GR), a strange form of energy density exhibiting high negative pressure, known as dark energy (DE), has been postulated to account for this phenomenon. Our Universe has $76\%$ dark energy according to the most recent CMBR data. Astronomers worry that as the universe expands, its matter may eventually be gradually split apart. Cosmologists introduces this phenomenon as Big Rip (BR). When time is finite, both the scale factor and phantom energy density arrive at infinity \cite{P03}. It should be noted that the phantom dark energy hypothesis is also supported by the available observational data. This model may help to resolve the Hubble tension problem \cite{S20}. The equation of state parameter $(\omega)$ is less than $-1$ in this case. Under such circumstances, the Universe's evolution would transition into a phantom phase, resulting in super-accelerated expansion and potentially leading to a future singularity. In conventional matter, the reduction in scale factor led to an increase in energy density, while  the energy density increased as the scale factor grew for the phantom energy-dominated stage. Growing energy densities are also the foundation for the Universe's expansion. The BR singularity occurs even in the presence of violations of all energy conditions. Thus, it is possible to categorise the phantom dark energy hypotheses as: $(a)$ energy density eventually reaches a constant value in a finite amount of time, known as the cosmological constant and $(b)$ phantom energy in the absence of BR singularity. Various characteristics of BR singularity are discussed in the literature \cite{A00,Nojiri03}. The requirement of an EoS value less than $-1$ does not seem to be adequate for the singularity to occurred. When time progresses, the energy density either increases or stays constant, and the EoS value may eventually approach $-1$. Therefore, the possibility of a finite-time future singularity may not arise \cite{P11}. Pseudo Rip (PR) could be associated with the Hubble rate when it approaches a constant value \cite{AV12}. Thus, in the PR, the energy density increases monotonically with the scale factor. The disintegration of bound structures may result from these kinds of models, given sufficiently powerful inertial forces \cite{KJ12}. The LR hypothesis is an additional form of rip cosmology where the energy density increases monotonically without an upper bound, potentially avoiding a future singularity \cite{R03}.
 
 A number of substitute hypotheses that do not include the cosmological constant have been postulated for comprehending the accelerating stage of the cosmos. These alternative theories of gravity, often known as modified gravity theories, are predicated on alterations to Einstein's General Relativity. Since it provides substantial answers to many fundamental concerns concerning dark energy and the late time accelerating Universe, modified gravity is an attractive possibility. DE models based on modified gravity can provide effective solutions that are consistent with recent observational data. Changes in the geometry of space-time have led to the discovery of models such as $f(R)$ gravity \cite{Nojiri11}, $f(G)$ is also called Gauss-Bonnet gravity where $G$ is the Gauss-Bonnet scalar \cite{Odintsov07}, Lovelock gravity \cite{T13}, scalar tensor theories \cite{Luca99}, braneworld models \cite{Gia00}, etc. Indeed, $f(Q)$ gravity, which originates from non-metricity, provides an alternative method for constructing various categories of modified gravities \cite{J99,Jimenez18}. The distinctive classes of modified gravity mentioned above arise from considerations of curvature, torsion, and non-metricity, even though the unmodified theories are equivalent at the level of equations. The reason for this distinction lies in the fact that unlike the standard Levi-Civita Ricci scalar $\overset{\circ}R$ in general relativity, the torsion scalar $T$ and the non-metricity scalar $Q$ contain a total divergence term, leading to equations such as $\overset{\circ}R=-T+B$ and $\overset{\circ}R=Q+C$. It means that the arbitrary functions $f(\overset{\circ}R)$, $f(T)$ and $f(Q)$ no longer differ by a complete derivative. The boundary B can be included into the Lagrangian in the context of teleparallel gravities to produce $f(T,B)$ theories, which naturally have wealthy phenomenon \cite{S23}. However, the Lagrangian of symmetric teleparallel gravity does not include the role of C within the paradigm of non-metricity. Cosmologists are currently interested in the $f(Q,C)$ theory that is thus developed \cite{De52,V0}. Modified gravity theories give a number of rip cosmological models. The second thermodynamics law is consistently satisfied for the LR model in $f(R,T)$ gravity, as demonstrated by \cite{MJS14}. Rip cosmological models in $f(T,B)$ gravity investigated by \cite{Rao24}. In anisotropic $f(R,T)$ gravity, the aspects of physical parameters investigated by \cite{B20}. Rip cosmological models in $f(Q,T)$ gravity analyzed by \cite{Pati22}. Inspired by this investigations, we analyze the LR, BR and PR models in $f(Q,C)$ gravity theory.
 
 The layout of the article is as follows: We present an illustration of the field equations in $f(Q,C)$ gravity in part \ref{sec2}. In section \ref{sec3}, we analyze the LR, BR and PR models in $f(Q,C)$ gravity theory and investigate the aspects of all the cosmological parameters.. The energy conditions are given in part \ref{sec4}. Conclusions are displayed in portion \ref{sec5}.
 
 \section{$f(Q,C)$ gravity and field equations}\label{sec2}
 \hspace{0.5cm} It is well known that the only affine connection that meets both the metric compatibility and torsion-free requirements is the Levi-Civita connection $(\overset{\circ}{\Gamma^{\eta}}_{\mu\nu})$. We remove this constraint and proceed to create the symmetric teleparallel gravity by assuming an affine connection $\Gamma^{\eta}_{\mu\nu}$ that is free of both curvature and torsion. The word ``symmetric" refers to the affine connection's symmetry in its lower exponents, which is caused by the torsion free circumstance. The non-metricity tensor indicates that this affine connection is not compatible with the metric which is described by
\begin{equation}\label{1}
 Q_{\alpha\mu\nu}=\nabla_{\alpha}g_{\mu\nu}=\partial_{\alpha}g_{\mu\nu}-\Gamma^{\beta}_{\alpha\mu}g_{\beta\nu}
 \Gamma^{\beta}_{\alpha\nu}g_{\beta\mu}\neq 0.
 \end{equation}
 where $\Gamma^{\alpha}_{\mu\nu}=\overset{\circ}{\Gamma^{\alpha}}_{\mu\nu}+L^{\alpha}_{\mu\nu}$. It implies that
  \begin{equation}\label{2}
 L^{\alpha}_{\mu\nu}=\frac{1}{2}(Q^{\alpha}_{\mu\nu}-Q_{\mu\;\;\nu}^{\;\;\alpha}-Q_{\nu\;\;\mu}^{\;\;\alpha}).
 \end{equation}
 Two distinct kinds of non-metricity vectors can be created as
 \begin{equation}\label{3}
 Q_{\mu}=g^{\nu\alpha}Q_{\mu\nu\alpha}=Q_{\mu\;\;\nu}^{\;\;\nu},\hspace{0.5cm} \tilde{Q}_{\mu}=g^{\nu\alpha}Q_{\nu\mu\alpha}=Q_{\nu\mu}^{\;\;\;\;\nu}.
 \end{equation}
 The superpotential tensor $P^{\alpha}_{\mu\nu}$ is defined by
 \begin{equation}\label{4}
 P^{\alpha}_{\mu\nu}=\frac{1}{4}\bigg[-2L^{\alpha}_{\mu\nu}+Q^{\alpha}g_{\mu\nu}-\tilde{Q}^{\alpha}g_{\mu\nu}
 -\delta^{\alpha}_{\mu} Q_{\nu}\bigg].
 \end{equation}
 The non-metricity scalar $Q$ is obtained as
 \begin{equation}\label{5}
 Q=Q_{\eta\beta\gamma}P^{\eta\beta\gamma}.
 \end{equation}
 One can simply derive the following relations by using the limitations that are free of torsion and curvature:
 \begin{equation}\label{6}
 \overset{\circ}R_{\mu\nu}+\overset{\circ}\nabla_{\eta}L^{\eta}_{\mu\nu}-\overset{\circ}\nabla_{\nu}\tilde{L}_{\mu}
 +\tilde{L}_{\eta}LL^{\eta}_{\mu\nu}-L_{\eta\beta\nu}L^{\beta\eta}_{\mu}=0,
 \end{equation}
 and
 \begin{equation}\label{7}
 \overset{\circ}R+\overset{\circ}\nabla_{\eta}(L^{\eta}-\tilde{L}^{\eta})-Q=0.
 \end{equation} 
 By using the above relations, the boundary term $(C)$ is defined by
 \begin{equation}\label{8}
 C=\overset{\circ}R-Q=-\overset{\circ}\nabla_{\eta}(Q^{\eta}-\tilde{Q}^{\eta}).
 \end{equation}
 where $Q^{\eta}-\tilde{Q}^{\eta}=L^{\eta}-\tilde{L}^{\eta}$.\\
 The gravitational action of the $f(Q,C)$ gravity theory is given by
 \begin{equation}\label{9}
 S=\; \int \bigg(\frac{1}{2k}f(Q,C)+\mathcal{L}_{m}\bigg)\sqrt{-g}d^{4}x,
 \end{equation}
  where $f(Q,C)$ describes the function of non-metricity scalar $Q$ and the boundary term $C$ and the matter lagrangian is represented by $\mathcal{L}_{m}$.\\
 The field equation can be obtained by varying the action of equation (\ref{9}) in relation to the metric as follows:
 \begin{align}\label{10}
&\kappa T_{\mu\nu}=-\frac{f}{2}g_{\mu\nu}+\frac{2}{\sqrt{-g}}\partial_{\alpha}\bigg(\sqrt{-g}f_{Q}P^{\alpha}_{\mu\nu}\bigg)
+\bigg(P_{\mu\eta\beta}Q_{\nu}^{\eta\beta}-2P_{\eta\beta\nu}Q^{\eta\beta}_{\mu}\bigg)f_{Q}\\\nonumber
&+\bigg(\frac{C}{2}g_{\mu\nu}-\overset{\circ}\nabla_{\mu}\overset{\circ}{\nabla_{\nu}}+g_{\mu\nu}\overset{\circ}{\nabla^{\eta}}\overset{\circ}\nabla_{\eta}-2P^{\alpha}_{\mu\nu}\partial_{\alpha}\bigg)f_{C},
 \end{align}
The covariant form is
\begin{align}\label{11}
&\kappa T_{\mu\nu}=-\frac{f}{2}g_{\mu\nu}+2P^{\alpha}_{\mu\nu}\nabla_{\alpha}(f_{Q}-f_{C})+\bigg(\overset\circ G_{\mu\nu}+\frac{Q}{2}g_{\mu\nu}\bigg)f_{Q}\\\nonumber
&+\bigg(\frac{C}{2}g_{\mu\nu}-\overset{\circ}\nabla_{\mu}\overset{\circ}\nabla_{\nu}+g_{\mu\nu}\overset{\circ}{\nabla^{\eta}}\overset{\circ}\nabla_{\eta}\bigg)f_{C}.
\end{align}
The effective energy momentum tensor is specified as follows:
\begin{align}\label{12}
&T_{\mu\nu}^{eff}=T_{\mu\nu}+\frac{1}{k}\bigg(\frac{f}{2}g_{\mu\nu}-2P^{\alpha}_{\mu\nu}\nabla_{\alpha}(f_{Q}-f_{C})
-\frac{Qf_{Q}}{2}g_{\mu\nu}-\bigg[\frac{C}{2}g_{\mu\nu}-\overset{\circ}\nabla_{\mu}\overset{\circ}\nabla_{\nu}+g_{\mu\nu}\overset{\circ}{\nabla^{\eta}}\overset{\circ}\nabla_{\eta}\bigg]f_{C}\bigg).
\end{align}
By using the above equation we generate an equation similar to GR as follows:
\begin{equation}\label{13}
\overset{\circ}G_{\mu\nu}=\frac{k}{f_{Q}}T_{\mu\nu}^{eff}.
\end{equation}
We assume the energy momentum tensor for the perfect fluid as
\begin{equation}\label{14}
T_{\mu\nu}=pg_{\mu\nu}+(\rho+p)u_{\mu}u_{\nu}.
\end{equation}
where $\rho$ indicates the energy density, $p$ indicates the pressure and $u^{\mu}$ represents the four velocity of the fluid. We assume a spatially flat, homogeneous, isotropic Friedmann–Robertson–Walker (FRW) model of the Universe, denoted by the line-element
\begin{equation}\label{15}
  ds^{2}=-dt^{2}+a^{2}(t)(dx^{2}+dy^{2}+dz^{2}),
\end{equation}
where the scale factor is denoted by $a(t)$. Using the disappearing affine connection $\Gamma^{\eta}_{\mu\nu}=0$, we calculate the necessary values are as follows:
\begin{equation}\label{16}
\overset{\circ}R=6(2H^{2}+\dot{H}), \hspace{0.7cm} Q=-6H^{2}, \hspace{0.7cm} C=6(3H^{2}+\dot{H}).
\end{equation}
The field equations are derived by using the above expressions of equation (\ref{16}) as follows:
\begin{equation}\label{17}
\kappa\rho=\frac{f}{2}+6H^{2}f_{Q}-(9H^{2}+3\dot{H})f_{C}+3H\dot{f}_{C},
\end{equation}
\begin{equation}\label{18}
\kappa p=-\frac{f}{2}-(6H^{2}+2\dot{H})f_{Q}-2H\dot{f}_{Q}+(9H^{2}+3\dot{H})f_{C}-\ddot{f}_{C}.
\end{equation}
where $f_{Q}=\frac{\partial f}{\partial Q}$, $f_{C}=\frac{\partial f}{\partial C}$ and the over dot represents the derivative w. r. t. the time $``t"$. To ascertain suitable solutions for the field equations, it's essential to make specific assumptions concerning cosmological models integrating both $Q$ and $C$. The form of the field equations (\ref{17}) and (\ref{18}) are non-linear, so to find the solutions of these equations are not so easy. We consider a non-linear form of $f(Q,C)$ gravity as
\begin{equation}\label{19}
f(Q,C)=\alpha Q^{n}+\beta C,
\end{equation}
where $\alpha$, $\beta$ and $n$ are constants. \cite{Rana24} assumed $f(Q)=-Q+\alpha Q^{n}$ form to analyse the dynamical system in $f(Q)$ gravity theory. \cite{Maurya24} consider $f(Q,B)=\alpha Q^{2}+\beta B^{2}$ form to investigate the quintessence behavior in $f(Q,B)$ gravity theory. Various non-linear forms were employed by multiple authors in different gravity theories. We take into account this particular non-linear form in our computation, which was inspired by the models mentioned above. By using the equation (\ref{19}) in equations (\ref{17}) $\&$ (\ref{18}), the expressions of the energy density and pressure are obtained in the following manner:
\begin{equation}\label{20}
\rho=-\alpha(-6)^{n}H^{2n}\bigg(n-\frac{1}{2}\bigg),
\end{equation}
\begin{equation}\label{21}
p=\alpha(-6)^{n}H^{2n}\bigg(n-\frac{1}{2}\bigg)+\frac{\alpha n}{3}\dot{H}(-6)^{n}H^{2n-2}(2n-1),
\end{equation}
We plan to investigate the potential emergence of a singularity in the coming time, whether it be within a finite timescale or extending infinitely into the future. Consequently, we've explored different expressions of the Hubble parameter to gain insight into the future trajectory of the Universe's evolution.
\section{The models}\label{sec3}
\hspace{0.5cm} In this portion, we will talk about three future singularity stages, namely Little Rip (LR), Big Rip (BR) and Pseudo Rip (PR), viewing them as distinct cosmological models. Since cosmological investigations have confirmed that the Universe is expanding, there is a possibility that the Universe will eventually explode as a result of phantom energy building up. We aim to present the possible outcomes of the theoretical events and what they mean for the Universe in the future. In order to investigate the geometrical and dynamical aspects of the rip models, the necessary parameters are as follows:
\begin{equation}\label{22}
q=-1-\frac{\dot{H}}{H^{2}}, \hspace{0.6cm} j=2q^{2}+q-\frac{\dot{q}}{H}, \hspace{0.6cm} s=\frac{r-1}{3\bigg(q-\frac{1}{2}\bigg)}.
\end{equation}
To provide a clearer understanding,, we will analyze the physical characteristics of the parameters in relation to redshift, which can be linked to the scale factor through the equation $a=\frac{1}{1+z}$.
\subsection{Little rip}\label{sec3.1}
\hspace{0.5cm} The scale factor of the LR model can be expressed as,
\begin{equation}\label{23}
a(t)=a_{0}e^{\bigg[\frac{A}{\lambda}(e^{t\lambda}-e^{t_{0}\lambda})\bigg]},
\end{equation}
where $A$ and $\lambda$ represent positive constants. The Hubble parameter is evaluated by using the above scale factor's form as $H=Ae^{\lambda t}$ and it exhibits LR behavior when $\lambda>0$. The current epoch's scale factor is designated as $a_{0}=1$. The expression of cosmic time $(t)$ is obtained by using the equation (\ref{23}) as,
\begin{equation}\label{24}
t(z)=\frac{1}{\lambda}\bigg[log\bigg(-\frac{\lambda}{A}log(1+z)+e^{t_{0}\lambda}\bigg)\bigg],
\end{equation}
The Hubble parameter's expression in terms of redshift can be derived by employing the $t-z$ relation from equation (\ref{24}), as follows:
\begin{equation}\label{25}
H(z)=-\lambda log(1+z)+Ae^{t_{0}\lambda}.
\end{equation}
\begin{figure}[h!]
\centering
  \includegraphics[scale=0.5]{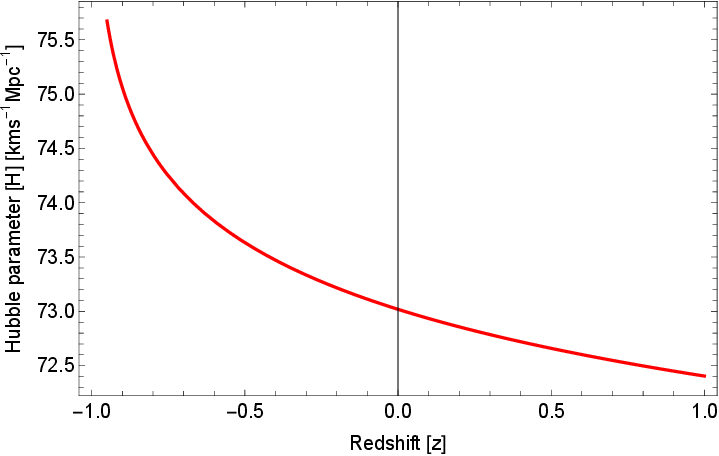}
  \caption{Plot of $H(z)$ vs $z$ for LR model with $A=21.31$, $\lambda=0.886$ and $t_{0}=1.39$.}
  \label{fig:f1}
\end{figure}

Figure \ref{fig:f1} illustrates the relationship between the Hubble parameter and redshift for the LR model. The figure demonstrates that the Hubble rate diverges as $t\rightarrow \infty$ and  it exhibits exponential growth over time. The current value of the Hubble parameter is determined as $H_{0}=73.03\; kms^{-1}Mpc^{-1}$ for $A=21.31$, $\lambda=0.886$ and $t_{0}=1.39$. With reference to equation (\ref{23}), we can compute the deceleration parameter in the following manner:
\begin{equation}\label{26}
q(t)=-1-\frac{\lambda}{Ae^{t\lambda}},
\end{equation}
\begin{figure}[h!]
\centering
  \includegraphics[scale=0.5]{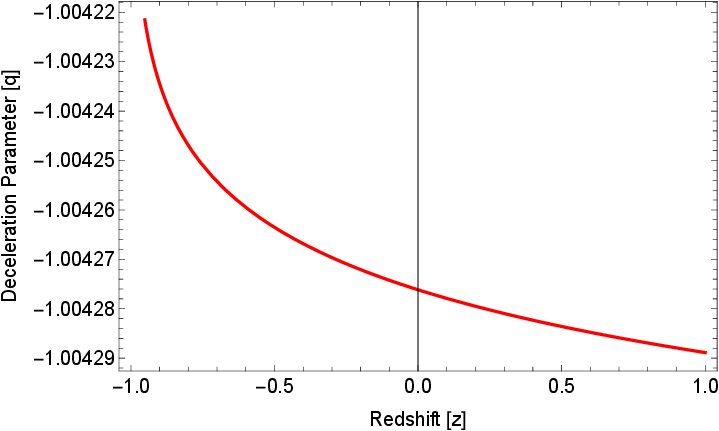}
  \caption{Plot of $q(z)$ vs $z$ for LR model with $A=21.31$, $\lambda=0.886$ and $t_{0}=1.39$.}
  \label{fig:f2}
\end{figure}

We can observe from equation (\ref{26}) that the deceleration parameter represents the negative behavior for any positive values of $A$ and $\lambda$. It transitions from a lower negative value to $-1$ during the late stages of evolution. As $e^{\lambda t}$ is always positive, the deceleration parameter remains negative for any positive values of $\lambda$, which depicts the Universe's accelerated stage. The deceleration parameter's expression in terms of redshift is derived from equation (\ref{24}), as follows:
\begin{equation}\label{27}
q(z)=-1-\frac{\lambda}{(-\lambda log(1+z)+Ae^{t_{0}\lambda})}.
\end{equation}
Figure \ref{fig:f2} illustrates the relationship between the deceleration parameter and redshift for the LR model. The figure displays only the accelerated epoch. The current value of the deceleration parameter is determined as $q_{0}=-1.0043$ which is similar to the observational value \cite{Marra20}. It is not possible to depict the transient behaviour of the deceleration parameter since the LR model prohibits singularity at finite time. The jerk and snap parameters for the LR model are as follows:
\begin{equation}\label{28}
j=1+\frac{\lambda^{2}}{A^{2}e^{2\lambda t}}+\frac{3\lambda}{Ae^{\lambda t}},
\end{equation} 
and
\begin{equation}\label{29}
s=\frac{\frac{\lambda^{2}}{A^{2}e^{2\lambda t}}+\frac{3\lambda}{Ae^{\lambda t}}}{3\bigg(-\frac{3}{2}-\frac{\lambda}{Ae^{\lambda t}}\bigg)}.
\end{equation}
\begin{figure}[hbt!]
    \centering
        \includegraphics[scale=0.5]{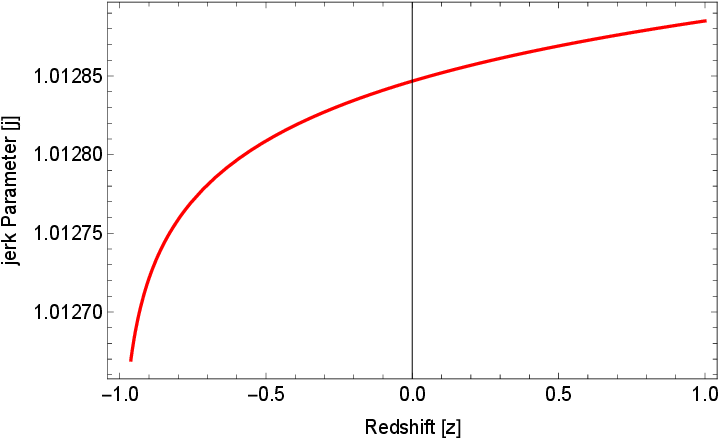}~~~~~~~~~~~
        \includegraphics[scale=0.5]{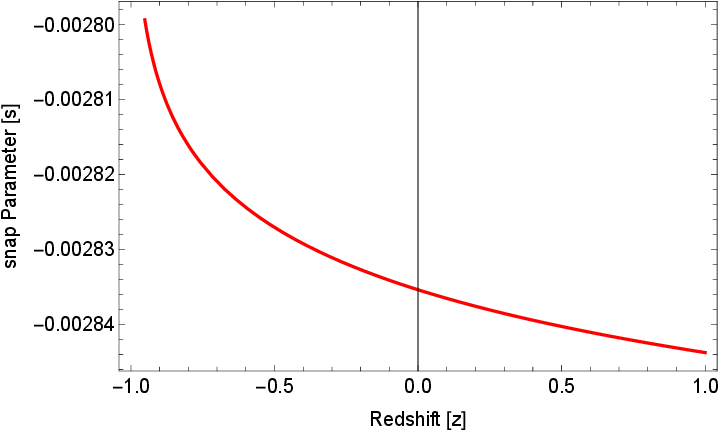}~~~~~~~~~~

    \caption{Plot of jerk and snap parameters vs $z$ for LR model with $A=21.31$, $\lambda=0.886$ and $t_{0}=1.39$.}
    \label{fig:f3}
\end{figure}

Figure \ref{fig:f3} illustrates the relationship between the jerk and snap parameters and redshift for the LR model. The nature of the jerk parameter is decreasing, while the snap parameter describes its increasing nature throughout the redshift $(z)$. The current values of the jerk and snap parameters are determined as $j_{0}=1.013$ and $s_{0}=-0.0028$. Our findings agree with the available observable data. The Universe's accelerated stage is described by the EoS parameter $(\omega)$. By inserting the form of $f(Q,C)$ from equation (\ref{19}) in the equations (\ref{20}) and (\ref{21}), the energy density and pressure for the LR model are described as follows:
\begin{equation}\label{30}
\rho=-\alpha(-6)^{n}\bigg(n-\frac{1}{2}\bigg)(-\lambda log(1+z)+Ae^{t_{0}\lambda})^{2n},
\end{equation}
\begin{equation}\label{31}
p=\alpha(-6)^{n}\bigg(n-\frac{1}{2}\bigg)(-\lambda log(1+z)+Ae^{t_{0}\lambda})^{2n}+\frac{\alpha n}{3}(-6)^{n}(2n-1)\lambda(-\lambda log(1+z)+Ae^{t_{0}\lambda})^{2n-1}.
\end{equation}
\begin{figure}[hbt!]
    \centering
        \includegraphics[scale=0.5]{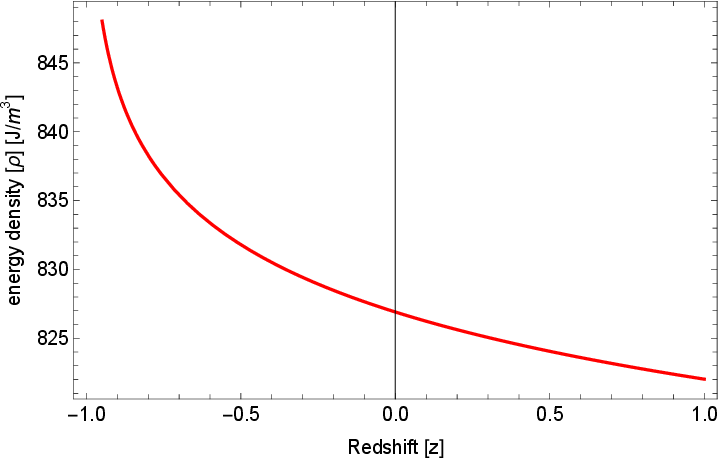}~~~~~~~~~~~
        \includegraphics[scale=0.5]{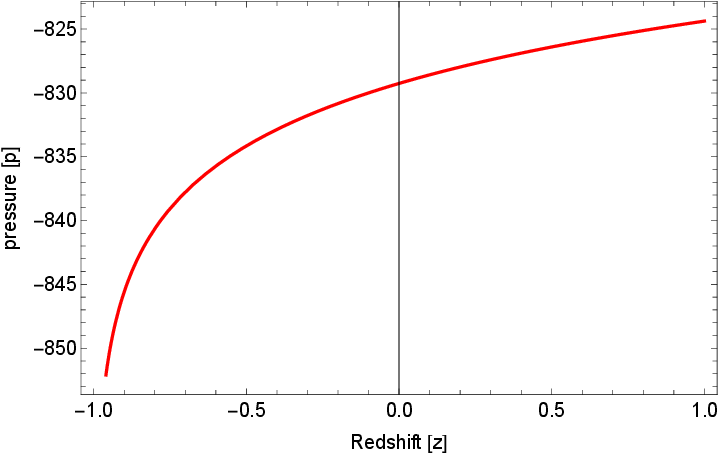}~~~~~~~~~~

    \caption{Plot of energy density and pressure vs $z$ for LR model with $A=21.31$, $\lambda=0.886$, $t_{0}=1.39$, $\alpha=0.5$ and $n=1$.}
    \label{fig:f4}
\end{figure}

The nature of the energy density and pressure are dependent upon the model parameters $A$, $t_{0}$, $\lambda$, $\alpha$ and $n$. In order to maintain the Hubble and deceleration parameters within the bounds indicated by cosmological findings, we choose the value of the parameter $\lambda=0.886$. Subsequently, we set the values of our model parameters $\alpha$ and $\beta$ accordingly to maintain a positive energy density and the EoS parameter's accelerating characteristics. In Figure \ref{fig:f4} (left plot), the energy density is consistently positive for all $z$ values and demonstrates an increasing behavior. In Figure \ref{fig:f4} (right plot), the pressure is consistently negative for all $z$ values and demonstrates an decreasing behavior. \\
The EoS parameter for LR is determined by utilising equations (\ref{30}) and (\ref{31}) in the following manner:
\begin{equation}\label{32}
\omega=\frac{p}{\rho}=-1-\frac{2n}{3}\frac{\lambda}{(-\lambda log(1+z)+Ae^{t_{0}\lambda})}.
\end{equation}
\begin{figure}[h!]
\centering
  \includegraphics[scale=0.5]{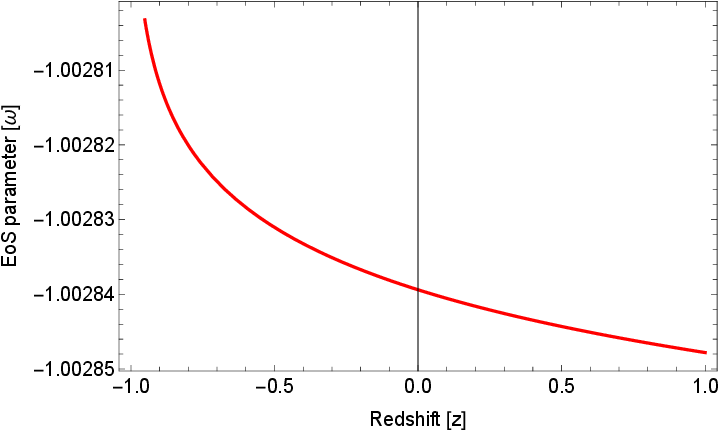}
  \caption{Plot of the EoS parameter ($\omega$) vs redshift ($z$) for LR with $A=21.31$, $\lambda=0.886$, $t_{0}=1.39$, $\alpha=0.5$ and $n=1$.}
  \label{fig:f5}
\end{figure}

In Figure \ref{fig:f5}, the graph illustrates how the LR's EoS parameter ($\omega$) changes concerning the redshift ($z$) with the suitable values of the model parameters. Dark energy properties and the age of the Universe are both elucidated by the EoS parameter. $\omega$ spans a range of values from $-1$ to $0$. For $\omega=1$, it denotes a stiff fluid; for $-1<\omega<-0.333$, it signifies the quintessence stage; $\omega=-1$ corresponds to the $\Lambda$CDM stage; and for $\omega>-1$, it indicates the phantom stage. The path in Figure \ref{fig:f5} reveals that the EoS parameter is negative for all $z$ values and initially, it belongs to the phantom stage and gradually approaches a significantly smaller value in the future. Additionally, the present value of $\omega$ is $\omega_{0}=-1.0028$. Our result is in agreement with the current observational data \cite{Silk16}.
\subsection{Big Rip}\label{sec3.2}
\hspace{0.5cm} A Big Rip occurs when the expansion of the Universe accelerates to such a high degree that it disrupts every formations, including stars, galaxies, atoms, and fundamental particles. Within a finite timeframe, the Universe's rate of the diverges towards infinity as the scale factor of the Universe undergoes exponential growth. The scale factor of the BR model can be expressed as,
\begin{equation}\label{33}
a(t)=a_{0}+\frac{1}{(t_{s}-t)^{\zeta}},
\end{equation}
where $t_{s}$ and $\zeta$ represent positive constants and $a_{0}$ denotes the constant of integration. In equation (\ref{33}), if $t=t_{s}$, then $a(t)$ approaches to infinity and if $``t"$ tends to infinity, then $a(t)$ becomes $a_{0}$. The Hubble parameter can be derived by using the equation (\ref{33}) as follows:
\begin{equation}\label{34}
H=\frac{\zeta}{(t_{s}-t)(a_{0}(t_{s}-t)^{\zeta}+1)}.
\end{equation}
We noticed that for $a_{0}=0$, $H(t)$ becomes $H=\frac{\zeta}{t_{s}-t}$. The expression of cosmic time $(t)$ is obtained by using the equation (\ref{23}) as,
\begin{equation}\label{35}
t(z)=t_{s}-\bigg(\frac{1+z}{1-a_{0}(1+z)}\bigg)^{\frac{1}{\zeta}},
\end{equation} 
\begin{figure}[h!]
\centering
  \includegraphics[scale=0.5]{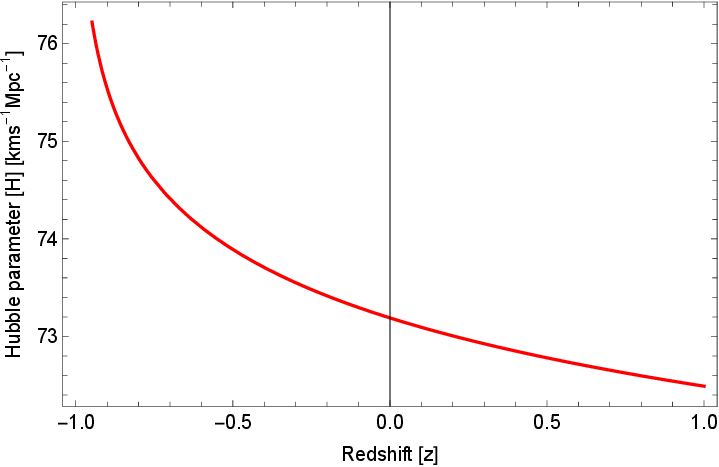}
  \caption{Plot of $H(z)$ vs $z$ for BR model with $\zeta=73.2$ and $a_{0}=0.01$.}
  \label{fig:f6}
\end{figure}

Figure \ref{fig:f6} illustrates the relationship between the Hubble parameter and redshift for the BR model. The Hubble parameter demonstrates the increasing behavior throughout $z$. The current value of the Hubble parameter is determined as $H_{0}=73.19\; kms^{-1}Mpc^{-1}$ for $\zeta=73.2$ and $a_{0}=0.01$. With reference to equation (\ref{34}), we can compute the deceleration parameter in the following manner:
\begin{equation}\label{36}
q(t)=-\frac{(\zeta+1)[1+a_{0}(t_{s}-t)^{\zeta}]}{\zeta},
\end{equation}
We noticed that for $\zeta<-1$, $q>0$ illustrates the decelerated stage, for $\zeta>-1$, $q<0$ illustrates the accelerated stage, while at $\zeta=-1$, $q=0$ illustrates the marginal expansion. The deceleration parameter's expression in terms of redshift is derived from equation (\ref{35}), as follows:
\begin{equation}\label{37}
q(z)=-\bigg(\frac{\zeta+1}{\zeta}\bigg)\bigg(\frac{1}{1-a_{0}(1+z)}\bigg).
\end{equation}
\begin{figure}[h!]
\centering
  \includegraphics[scale=0.5]{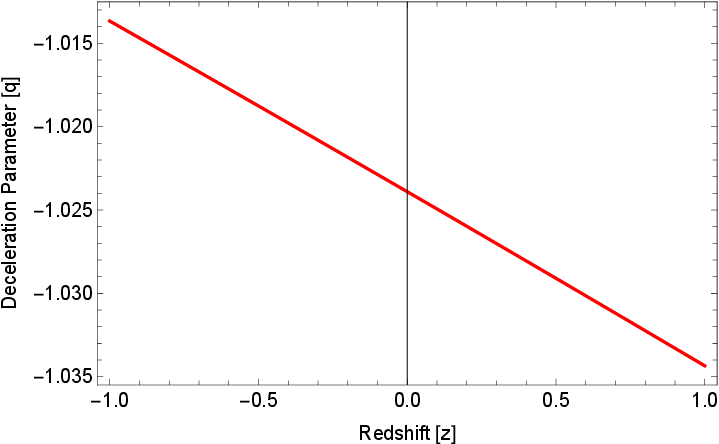}
  \caption{Plot of $q(z)$ vs $z$ for BR model with $\zeta=73.2$ and $a_{0}=0.01$.}
  \label{fig:f7}
\end{figure}

Figure \ref{fig:f7} illustrates the relationship between the deceleration parameter and redshift for the BR model. The figure displays the increasing behavior. The current value of the deceleration parameter is determined as $q_{0}=-1.024$ for $\zeta=73.2$ and $a_{0}=0.01$ represents the Universe's accelerated expansion. The jerk and snap parameters for the BR model are as follows:
\begin{equation}\label{38}
j(z)=\bigg(\frac{(\zeta+1)(\zeta+2)}{\zeta^{2}}\bigg)\bigg(\frac{1}{1-a_{0}(1+z)}\bigg)^{2},
\end{equation}
\begin{equation}\label{39}
s(z)=\bigg(\frac{(\zeta+1)(\zeta+2)(\zeta+3)}{\zeta^{3}}\bigg)\bigg(\frac{1}{1-a_{0}(1+z)}\bigg)^{3}.
\end{equation}
\begin{figure}[hbt!]
    \centering
        \includegraphics[scale=0.5]{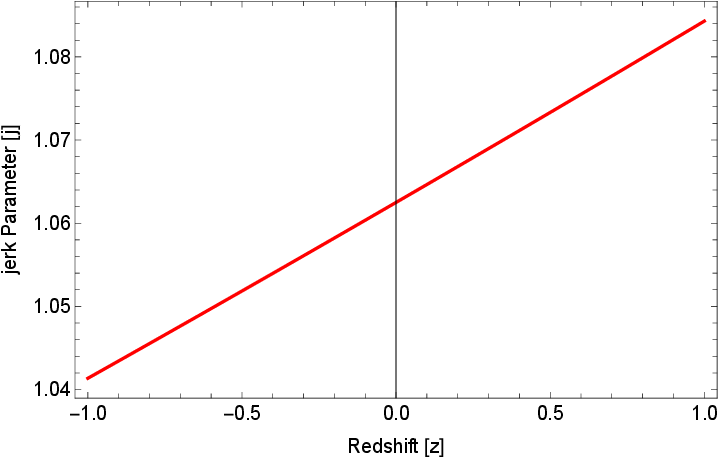}~~~~~~~~~~~
        \includegraphics[scale=0.5]{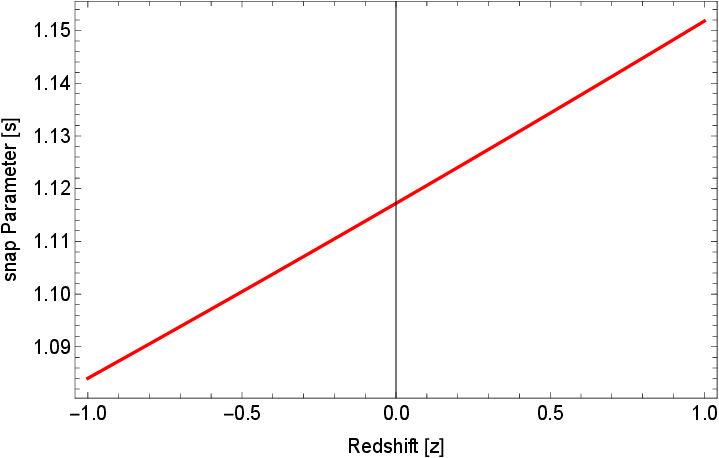}~~~~~~~~~~

    \caption{Plot of jerk and snap parameters vs $z$ for BR model with $\zeta=73.2$ and $a_{0}=0.01$.}
    \label{fig:f8}
\end{figure}

Figure \ref{fig:f8} illustrates the relationship between the jerk and snap parameters and redshift for the BR model. The nature of the jerk and snap parameters are decreasing nature throughout the redshift $(z)$. The current values of the jerk and snap parameters are determined as $j_{0}=1.0624$ and $s_{0}=1.1171$. Our findings agree with the available observable data. The form of the energy density and pressure for the BR model are derived as follows:
\begin{equation}\label{40}
\rho=-\alpha(-6)^{n}\bigg(n-\frac{1}{2}\bigg)\zeta^{2n}\bigg(\frac{1-a_{0}(1+z)}{1+z}\bigg)^{\frac{2n}{\zeta}},
\end{equation}
\begin{equation}\label{41}
p=\alpha(-6)^{n}\bigg(n-\frac{1}{2}\bigg)\zeta^{2n}\bigg(\frac{1-a_{0}(1+z)}{1+z}\bigg)^{\frac{2n}{\zeta}}+\frac{\alpha n}{3}(-6)^{n}(2n-1)\zeta^{2n-1}\bigg(\frac{1-a_{0}(1+z)}{1+z}\bigg)^{\frac{2n}{\zeta}}.
\end{equation}
\begin{figure}[hbt!]
    \centering
        \includegraphics[scale=0.5]{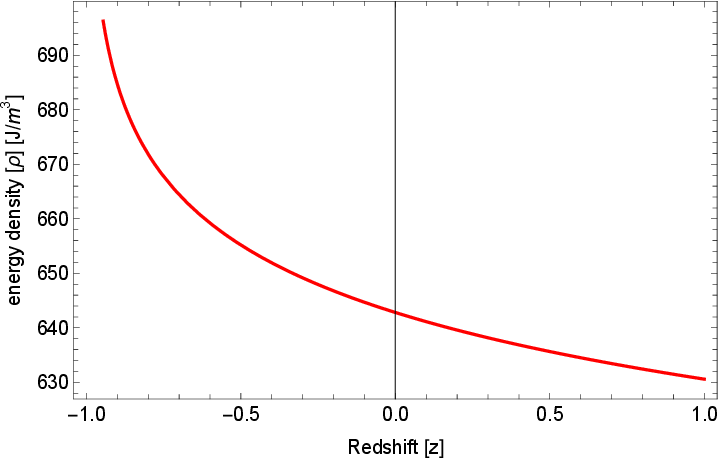}~~~~~~~~~~~
        \includegraphics[scale=0.5]{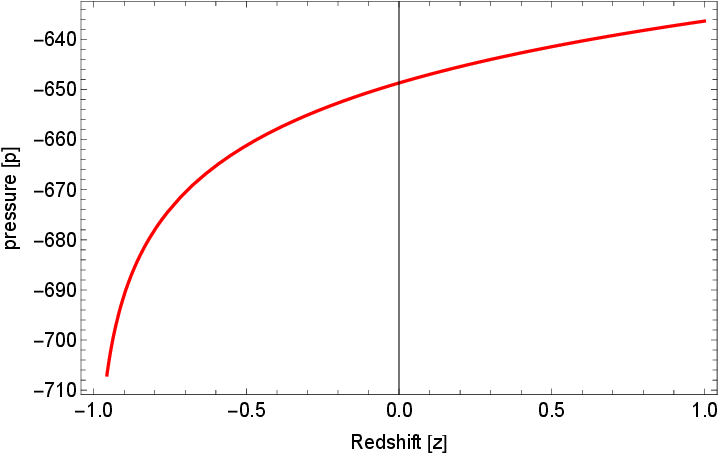}~~~~~~~~~~

    \caption{Plot of energy density and pressure vs $z$ for BR model with $\zeta=73.2$ and $a_{0}=0.01$, $\alpha=0.5$ and $n=1$.}
    \label{fig:f9}
\end{figure}

The nature of the energy density and pressure are dependent upon the model parameters $\zeta$, $a_{0}$, $\alpha$ and $n$. In order to maintain the Hubble and deceleration parameters within the bounds indicated by cosmological findings, we choose the value of the parameter $\zeta=73.2$. Subsequently, we set the values of our model parameters $\alpha$ and $\beta$ accordingly to maintain a positive energy density and the EoS parameter's accelerating characteristics. In Figure \ref{fig:f9} (left plot), the energy density is consistently positive for all $z$ values and demonstrates an increasing behavior. In Figure \ref{fig:f9} (right plot), the pressure is consistently negative for all $z$ values and demonstrates an decreasing behavior. \\
The EoS parameter for BR is determined by utilising equations (\ref{40}) and (\ref{41}) in the following manner:
\begin{equation}\label{42}
\omega=\frac{p}{\rho}=-1-\frac{2n}{3\zeta}\bigg[\frac{1+\zeta a_{0}(1+z)}{1-a_{0}(1+z)}\bigg].
\end{equation}
\begin{figure}[h!]
\centering
  \includegraphics[scale=0.5]{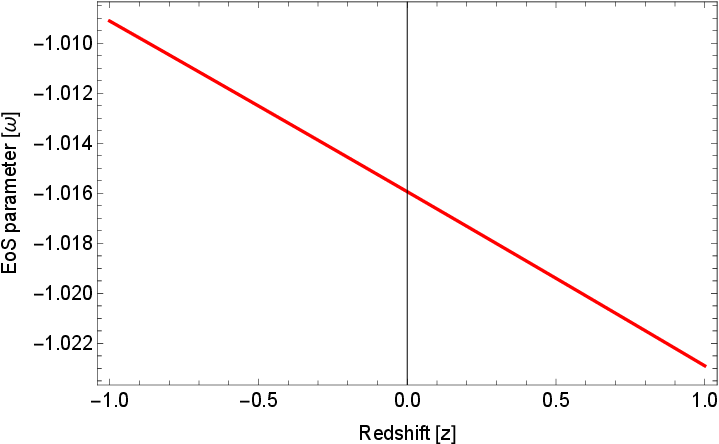}
  \caption{Plot of the EoS parameter ($\omega$) vs redshift ($z$) for BR with $\zeta=73.2$ and $a_{0}=0.01$.}
  \label{fig:f10}
\end{figure}

In Figure \ref{fig:f10}, the graph illustrates how the BR's EoS parameter ($\omega$) changes concerning the redshift ($z$) with the suitable values of the model parameters. The evolution of the Equation of State (EoS) parameter curve initially enters the phantom stage, steadily increases and subsequently remains within that region. In the late stages, it tends to approach the $\Lambda$CDM line. The present value of $\omega$ is $\omega_{0}=-1.0159$. Our result is in agreement with the current observational data.
\subsection{Pseudo Rip}\label{sec3.3}
\hspace{0.5cm} The PR stage lies between the BR and LR stages. Relative to the BR stage, the acceleration of the Universe's expansion is less rapid. However, the scale factor's evolution and the deviation in the rate of expansion occurs more slowly. In certain instances, the Universe seems to be moving towards behavior resembling a rip, but the consequences are not as extreme as in a BR stage. The scale factor of the PR model can be expressed as,
\begin{equation}\label{43}
a(t)=a_{0}e^{H_{0}t+\frac{H_{1}}{\epsilon e^{\epsilon t}}},
\end{equation}
where $H_{0}$, $H_{1}$ and $\epsilon$ are the positive constants. The Hubble parameter can be derived by using the equation (\ref{43}) as follows:
\begin{equation}\label{44}
H=H_{0}-\frac{H_{1}}{e^{\epsilon t}},
\end{equation}
\begin{figure}[h!]
\centering
  \includegraphics[scale=0.5]{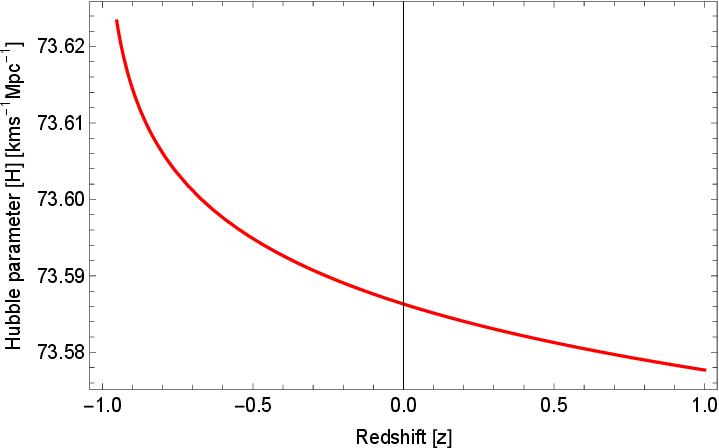}
  \caption{Plot of $H(z)$ vs $z$ for PR model with $H_{0}=74.6$, $H_{1}=1$ and $\epsilon=0.9$.}
  \label{fig:f11}
\end{figure}

It's worth noting that as $t\rightarrow \infty$, the Hubble parameter increases slowly with time, and it approaches the current value of the Hubble parameter \cite{N20}. Nevertheless, the Hubble parameter's higher order derivatives is discrete for this particular selection. Even if the Hubble parameter increases to an infinite value over time, throughout cosmological development, the inertial force dissociates a few of the model's bonded structures. Additionally, as $t \rightarrow \infty$, the scale factor approaches the de Sitter solution. Figure \ref{fig:f11} illustrates the relationship between the Hubble parameter and redshift for the PR model. The Hubble parameter demonstrates the increasing behavior throughout $z$. The current value of the Hubble parameter is determined as $H_{0}=73.586\; kms^{-1}Mpc^{-1}$ for $H_{0}=74.6$, $H_{1}=1$ and $\epsilon=0.9$. With reference to equation (\ref{44}), we can compute the deceleration parameter in the following manner:
\begin{equation}\label{45}
q=-1-\frac{\epsilon H_{1}e^{-\epsilon t}}{(H_{0}-H_{1}e^{-\epsilon t})^{2}}.
\end{equation}
\begin{figure}[h!]
\centering
  \includegraphics[scale=0.5]{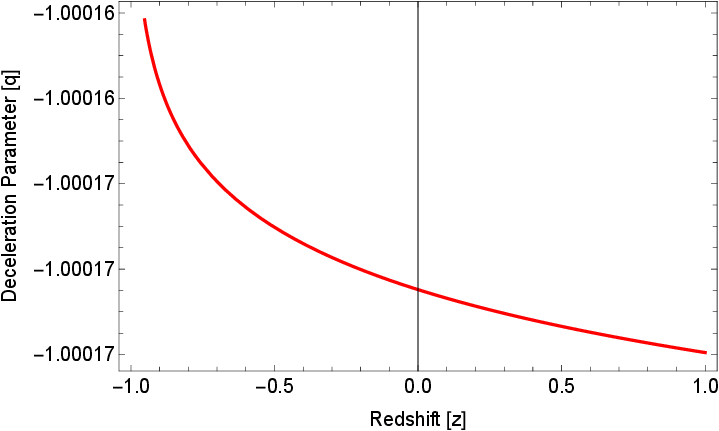}
  \caption{Plot of $q(z)$ vs $z$ for PR model with $H_{0}=74.6$, $H_{1}=1$ and $\epsilon=0.9$.}
  \label{fig:f12}
\end{figure}

In equation (\ref{45}), if $t= 0$ then $q=-1-\frac{\epsilon H_{1}}{H_{0}^{2}}$ and if $t\rightarrow \infty$ then $q\rightarrow -1$. Fixing the deceleration parameter's current value within the current observational data required us to set the parameters $H_{1}>0$ and $\epsilon> 0$. Figure \ref{fig:f12} illustrates the relationship between the deceleration parameter and redshift for the PR model. The deceleration parameter demonstrates the increasing behavior throughout $z$ in the negative portion. The current value of the deceleration parameter is determined as $q_{0}=-1.00017$ for $H_{0}=74.6$, $H_{1}=1$ and $\epsilon=0.9$. The jerk and snap parameters for the PR model are as follows:
\begin{equation}\label{46}
j=1-\frac{\epsilon e^{-2\epsilon t}H_{1}[e^{\epsilon t}(\epsilon-3H_{0})+3H_{1}]}{(H_{0}-H_{1}e^{-\epsilon t})^{3}}
\end{equation}
\begin{equation}\label{47}
s=\frac{e^{-3\epsilon t}\bigg(H_{0}^{4}e^{3\epsilon t}+e^{\epsilon t}H_{1}^{2}(6H_{0}^{2}-12H_{0}\epsilon+7\epsilon^{2})-e^{2\epsilon t}H_{1}(2H_{0}-\epsilon)(2H_{0}^{2}-2H_{0}\epsilon+\epsilon^{2})-2H_{1}^{3}(2H_{0}-3\epsilon)+e^{-\epsilon t}H_{1}^{4}\bigg)}{(H_{0}-H_{1}e^{-\epsilon t})^{4}}
\end{equation}
\begin{figure}[hbt!]
    \centering
        \includegraphics[scale=0.5]{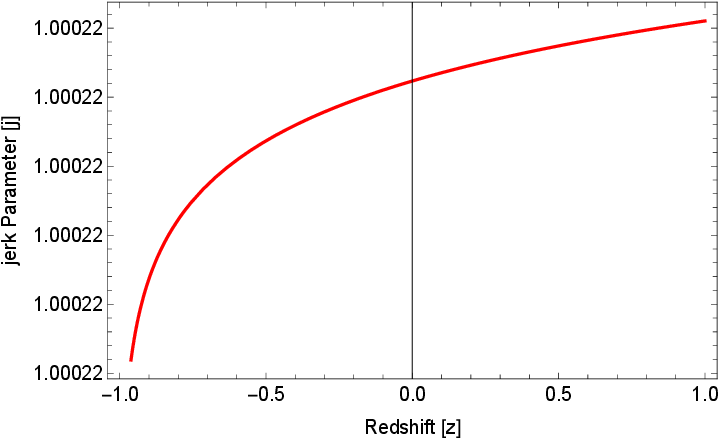}~~~~~~~~~~~
        \includegraphics[scale=0.5]{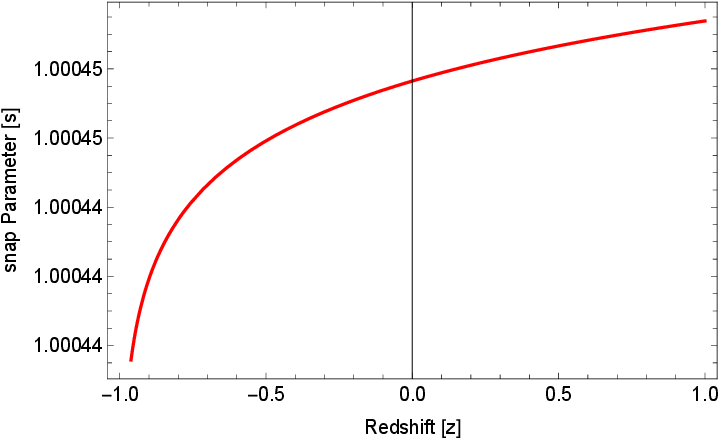}~~~~~~~~~~

    \caption{Plot of jerk and snap parameters vs $z$ for PR model with $H_{0}=74.6$, $H_{1}=1$ and $\epsilon=0.9$.}
    \label{fig:f13}
\end{figure}

Figure \ref{fig:f13} illustrates the relationship between the jerk and snap parameters and redshift for the PR model. The nature of the jerk and snap parameters are decreasing nature throughout the redshift $(z)$. The current values of the jerk and snap parameters are determined as $j_{0}=1.00022$ and $s_{0}=1.00045$. Our findings agree with the available observable data. The form of the energy density and pressure for the PR model are derived as follows:
\begin{equation}\label{48}
\rho=-\alpha(-6)^{n}\bigg(n-\frac{1}{2}\bigg)(H_{0}-H_{1}e^{-\epsilon t})^{2n},
\end{equation}
\begin{equation}\label{49}
p=\alpha(-6)^{n}\bigg(n-\frac{1}{2}\bigg)(H_{0}-H_{1}e^{-\epsilon t})^{2n}+\frac{\alpha n}{3}(-6)^{n}(2n-1)\epsilon H_{1}e^{-\epsilon t}(H_{0}-H_{1}e^{-\epsilon t})^{2n-2}.
\end{equation}
\begin{figure}[hbt!]
    \centering
        \includegraphics[scale=0.5]{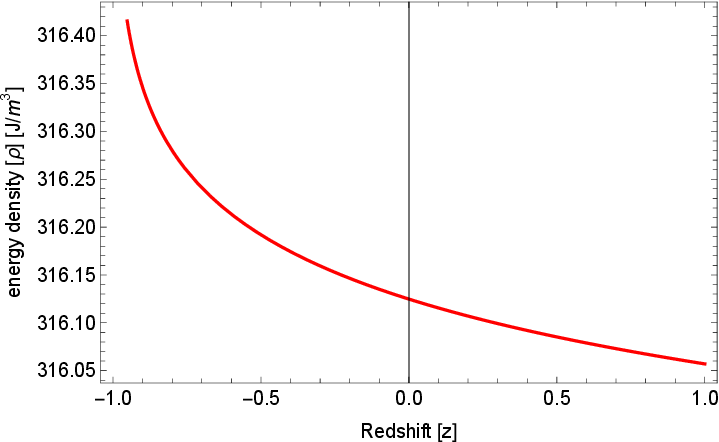}~~~~~~~~~~~
        \includegraphics[scale=0.5]{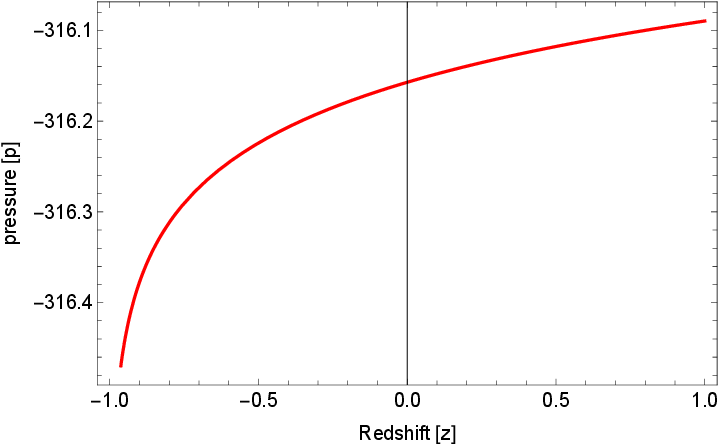}~~~~~~~~~~

    \caption{Plot of energy density and pressure vs $z$ for PR model with $H_{0}=74.6$, $H_{1}=1$ and $\epsilon=0.9$.}
    \label{fig:f14}
\end{figure}

In Figure \ref{fig:f14} (left plot), the energy density is consistently positive for all $z$ values and demonstrates an increasing behavior. In Figure \ref{fig:f14} (right plot), the pressure is consistently negative for all $z$ values and demonstrates an decreasing behavior. The EoS parameter for PR is determined by utilising equations (\ref{48}) and (\ref{49}) in the following manner:
\begin{equation}\label{50}
\omega=\frac{p}{\rho}=-1-\frac{2}{3}\frac{\epsilon H_{1}e^{-\epsilon t}}{(H_{0}-H_{1}e^{-\epsilon t})^{2}}.
\end{equation}
\begin{figure}[h!]
\centering
  \includegraphics[scale=0.5]{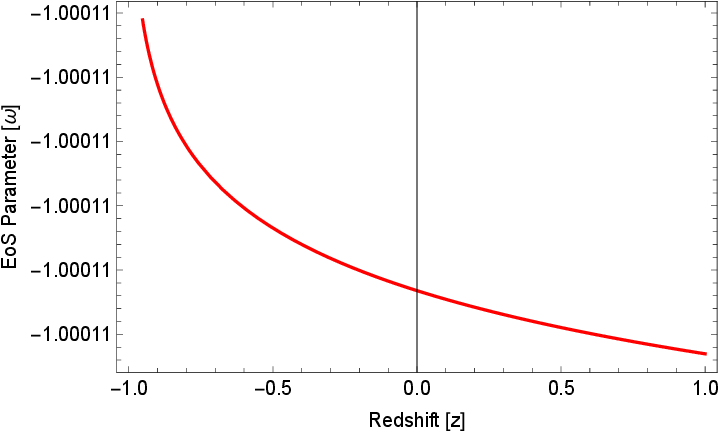}
  \caption{Plot of the EoS parameter ($\omega$) vs redshift ($z$) for PR with $H_{0}=74.6$, $H_{1}=1$ and $\epsilon=0.9$.}
  \label{fig:f15}
\end{figure}

In Figure \ref{fig:f15}, the graph illustrates how the PR's EoS parameter ($\omega$) changes concerning the redshift ($z$) with the suitable values of the model parameters. The evolution of the Equation of State (EoS) parameter curve decreases with time and it tends to approach the $\Lambda$CDM line. The present value of $\omega$ is $\omega_{0}=-1.00011$, which is near the $\Lambda$CDM model. Our result is in agreement with the current observational data.
\section{Energy Conditions}\label{sec4}
\hspace{0.5cm} In modified gravity theories, the central focus of theoretical modeling of the Universe concerns the viability of the models. In general relativity, distinct energy conditions provide restrictions on areas that cannot have a negative energy density. General Relativity's energy requirements are put to use for exploring black holes and spacetime singularity issues. \cite{Wald84}. The term ``energy conditions" encompasses the restrictions that define the correlation between pressure and energy density. The Raychaudhuri equations from 1995 can be used to determine the energy conditions \cite{Ray55}. The energy circumstances are categorized as \cite{Lobo15},\\
$\bullet$ Null Energy Conditions (NEC): $\rho+p\geq 0$,\\
$\bullet$ Weak Energy Conditions (WEC): $\rho\geq 0$, $\rho+p\geq 0$,\\
$\bullet$ Dominant Energy Conditions (DEC): $\rho-p\geq 0$,\\
$\bullet$ Strong Energy Conditions (SEC): $\rho+3p\geq 0$.\\
\begin{figure}[hbt!]
    \centering
    \begin{subfigure}[b]{0.3\textwidth}
        \includegraphics[scale=0.4]{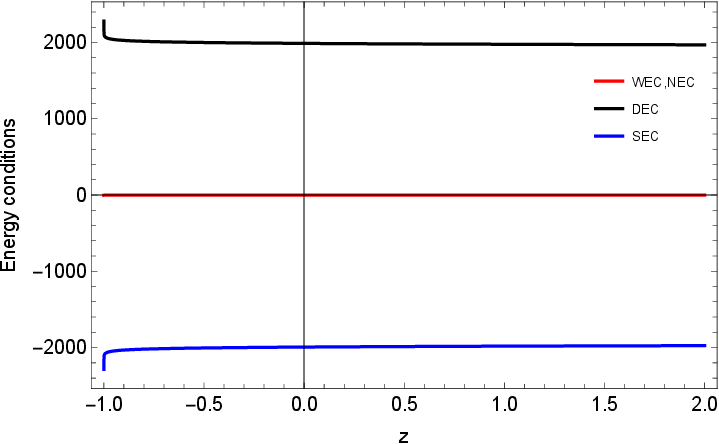}
    \end{subfigure}
    \hfill
    \begin{subfigure}[b]{0.3\textwidth}
        \includegraphics[scale=0.4]{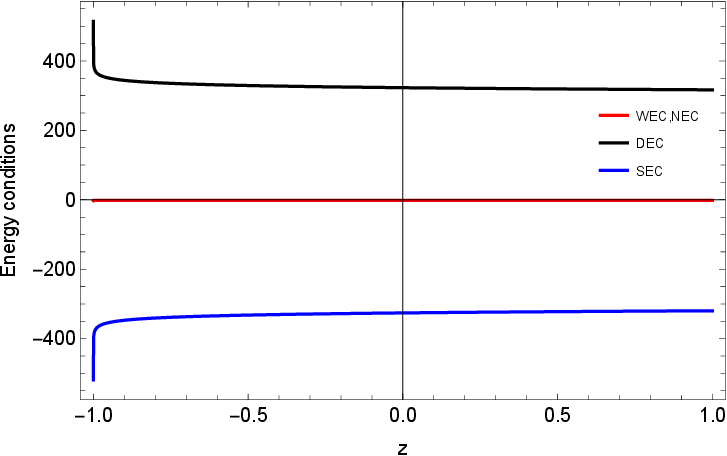}
    \end{subfigure}
    \hfill
    \begin{subfigure}[b]{0.3\textwidth}
        \includegraphics[scale=0.4]{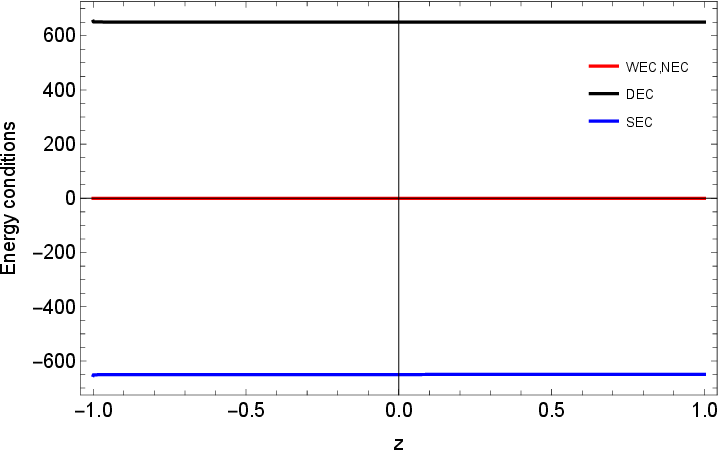}
    \end{subfigure}

    \caption{Behavior of energy conditions vs redshift for LR, BR and PR models.}
    \label{fig:f16}
\end{figure}\\

The LR models's energy conditions are derived by using the equations (\ref{30}) and (\ref{31}) as follows:
\begin{equation}\label{51}
\rho+p=\frac{\alpha n}{3}(-6)^{n}(2n-1)\lambda(Ae^{t\lambda})^{2n-1},
\end{equation}
\begin{equation}\label{52}
\rho-p=-2\alpha(-6)^{n}\bigg(n-\frac{1}{2}\bigg)(Ae^{t\lambda})^{2n}-\frac{\alpha n}{3}(-6)^{n}(2n-1)\lambda(Ae^{t\lambda})^{2n-1},
\end{equation}
\begin{equation}\label{53}
\rho+3p=2\alpha(-6)^{n}\bigg(n-\frac{1}{2}\bigg)(Ae^{t\lambda})^{2n}+\alpha n(-6)^{n}(2n-1)\lambda(Ae^{t\lambda})^{2n-1}.
\end{equation}
The BR models's energy conditions are derived by using the equations (\ref{40}) and (\ref{41}) as follows:
\begin{equation}\label{54}
\rho+p=\frac{\alpha n}{3}(-6)^{n}(2n-1)\zeta^{2n-1}\bigg(\frac{1}{t_{s}-t}\bigg)^{2n},
\end{equation}
\begin{equation}\label{55}
\rho-p=-2\alpha(-6)^{n}\bigg(n-\frac{1}{2}\bigg)\bigg(\frac{\zeta}{t_{s}-t}\bigg)^{2n}-\frac{\alpha n}{3}(-6)^{n}(2n-1)\zeta^{2n-1}\bigg(\frac{1}{t_{s}-t}\bigg)^{2n},
\end{equation}
\begin{equation}\label{56}
\rho+3p=2\alpha(-6)^{n}\bigg(n-\frac{1}{2}\bigg)\bigg(\frac{\zeta}{t_{s}-t}\bigg)^{2n}+\alpha n(-6)^{n}(2n-1)\zeta^{2n-1}\bigg(\frac{1}{t_{s}-t}\bigg)^{2n}
\end{equation}
The PR models's energy conditions are derived by using the equations (\ref{48}) and (\ref{49}) as follows:
\begin{equation}\label{57}
\rho+p=\frac{\alpha n}{3}(-6)^{n}(2n-1)\epsilon H_{1}e^{-\epsilon t}(H_{0}-H_{1}e^{-\epsilon t})^{2n-2},
\end{equation}
\begin{equation}\label{58}
\rho-p=-2\alpha(-6)^{n}\bigg(n-\frac{1}{2}\bigg)(H_{0}-H_{1}e^{-\epsilon t})^{2n}-\frac{\alpha n}{3}(-6)^{n}(2n-1)\epsilon H_{1}e^{-\epsilon t}(H_{0}-H_{1}e^{-\epsilon t})^{2n-2},
\end{equation}
\begin{equation}\label{59}
\rho+3p=2\alpha(-6)^{n}\bigg(n-\frac{1}{2}\bigg)(H_{0}-H_{1}e^{-\epsilon t})^{2n}+\alpha n(-6)^{n}(2n-1)\epsilon H_{1}e^{-\epsilon t}(H_{0}-H_{1}e^{-\epsilon t})^{2n-2}
\end{equation}

With the exception of the DEC, all models progress through the phantom phase and it is predicted that all other energy conditions will be violated. The energy conditions' characteristics for LR, BR and PR are depicted in figure \ref{fig:f16}. Across all models, the DEC is upheld within the appropriate range; while, as anticipated, both the NEC and SEC are violated. To enhance clarity and illustrate the violation of the NEC, it is incorporated into the figures. The NEC declined and continued to decrease, reaching a negative value for the negative region of the time. Therefore, the violated SEC coincides with the Universe's acceleration phase.
\section{Conclusion}\label{sec5}
\hspace{0.5cm} It is widely recognized that the problem of late-time cosmic acceleration has not been resolved within the framework of general relativity. But there is some cosmological understanding to be gained from the geometrically modified gravity. A further problem facing contemporary cosmology as a result of this cosmos's accelerated expansion is determining whether the Universe will eventually come to an end or remain intact indefinitely. Therefore, there is incentive to explore theoretical models concerning the rip singularity within the framework of $f(Q,C)$ gravity. Initially, we have determined the Hubble parameter's present value for the models within the range recommended by cosmic measurements. Next, we investigate the characteristics of the deceleration parameter. Avoiding a finite-time future singularity emerges as a key focus in cosmic investigations. The possibility that non-metricity, rather than the typical Ricci scalar, plays a part in preventing a finite-duration future singularity in this situation cannot be completely discounted. In each case, the deceleration parameter illustrates the Universe's accelerated expansion. We also investigate the behavior of the jerk and snap parameters in each case. For both LR and PR models, these parameters gradually decline with time, with a sharp decline occurring towards the end.

In the LR model, the EoS parameter exhibits a phantom nature, yet it stays extremely close to the $\Lambda$CDM line while consistently aligning with the $\Lambda$CDM line for the BR model. Meanwhile, in the PR model, it exhibits identical nature to that of the LR model, with the exception that it stays within an even smaller range. The current values of the EoS parameter for each model are $\omega_{0}=-1.0028, -1.0159, -1.00011$, respectively, compared to observational data \cite{R10}. Again, we also investigated the energy conditions and found that the SEC is violated and DEC satisfies the criteria for each model. But in each case, NEC belongs to the zero line or is displayed slightly below the zero line. It suggests that NEC's involvement in these models is essentially nonexistent. Our investigation leads us to the conclusion that the accelerating models do not contain a singularity situation. A more thorough examination of the $f(Q,C)$ models could provide fresh information on how to handle the singularity problem.

 \end{document}